\documentstyle[12pt]{article}

\evensidemargin = \oddsidemargin

\title{Is There a Problem with Quantum Gravitation?}
\author{Ranjan Kumar Ghosh\thanks{rkg\_1978@yahoo.com}\\
Bidhannagar College\\
EB-2, Sector I, Salt Lake City\\
Calcutta 700 064, INDIA}

\date{July 2002}

\begin{document}
\maketitle

\begin{abstract}
We argue that Quantum Gravitation forces us to sum over
metrics of all signatures.
\end{abstract}

Since the beginning of General Theory of Relativity the metric has
played the dynamical role of the gravitational potential. This has led
to an enormously successful classical theory of gravitation. All
subsequent efforts at a quantum theory of gravitation have therefore
been built around the idea that the graviton is a part of the
space-time metric.

In this short note we draw the attention towards a usually neglected
fact that a usual path integral quantization of the gravitational
field perhaps forces us to sum over space-time metrics of all
signatures. The signature of space-time metric is defined by the
positive and negative eigenvalues of the diagonalized metric. The
usual way of quantization makes us choose a ``classical'' background
metric of a certain signature (usually Lorentzian or Euclidean as the
case may be). We then consider the fluctuations given by all symmetric
matrices of the same dimensionality as the dimensionality of the
space-time under consideration. Actually we must divide the space of
all symmetric matrices by the group of General Coordinate
Transformations (GCT) as many of these matrices are related by GCT and
do not describe metrics differing from each other. However the GCT do
not change the signature of the metric. Therefore metrics with
different signatures correspond to different points in the quotient
space. It is actually the quotient space over which we have to perform
the sum.

The problem becomes obvious if we consider the simplest case of a
two-dimensional world. Let us consider the Euclidean case. In a
certain coordinate system the background metric takes the form
\begin{equation}
 \left( \begin{array}{cc}
 a & 0 \\
 0 & b 
\end{array} \right) 
\end{equation}
where $a$ and $b$ are both positive. Let us consider only those
quantum fluctuations of the gravitational field in which only the
diagonal elements are involved. Such a fluctuation is again
parametrized by a matrix
\begin{equation}
\left( \begin{array}{cc}
 c & 0 \\
 0 & d
\end{array} \right)
\end{equation}
For arbitrary values of $c$ and $d$ it is clear that the signature of
the sum of these two matrices is different from the original
background metric.  We may think that we may remedy the situation by
demanding that the magnitude of the larger of $c$ and $d$ be smaller
than the smaller of $a$ and $b$. However it is obviously
unnatural. Also just constraining the diagonal elements of
gravitational fluctuations does not solve the problem as the following
example shows. Let us take the starting background metric to be
\begin{equation}
\left( \begin{array}{cc}
 1 & 0 \\
 0 & 1
\end{array} \right)
\end{equation}
A fluctuation in the off-diagonal element puts this in a form\\
\begin{equation}
\left( \begin{array}{cc}
 1 & e \\
 e & 1
\end{array} \right) 
\end{equation}
When diagonalized by an orthogonal transformation, this becomes
\begin{equation}
\left( \begin{array}{cc}
 1+e & 0 \\
 0 & 1-e
\end{array} \right)
\end{equation}
Hence for $e>1$ the signature of this metric is Lorentzian.

The qualitative features remain the same in higher dimensions although
the calculations involved are more complicated.

This situation is however not completely new. Gibbons, Hawking and
Perry \cite{GHP} in their discussion of unboundedness of Euclidean
gravitational action came to the conclusion that in order to define
the path integral properly one needs to integrate along a contour in
the space of all metrics.

Hence from our discussion till now it is clear that if we are summing
over all graviton fluctuations around some background we are actually
summing over metrics of all signatures. If we constrain these
fluctuations somehow so that the summation extends over only one
signature of the metric, the usual expressions for the amplitudes may
not be obtained (those which we obtain in the usual field theories
from unrestricted summations).

Obviously there is the question of physical interpretation of these
contributing space-time metrics of different signatures.These will
naturally cause a violation of the Principle of Equivalence at the
quantum level (some metrics not having the correct signature). It is
at present under investigation. There may even be a connection between
this problem and the problem of smallness of the observed cosmological
constant.

These questions will hopefully be taken up in a future communication.

The author thanks Sunil Mukhi for illuminating discussions. He is also
thankful to Palash B. Pal for making Ref.~\cite{GHP} available and
suggesting a title (not the present one) for this note. I also thank
Rajkumar Chakraborty for helping me with \LaTeX\ for typing this note.

This work was supported partly by a UGC grant for a minor project
(sanction letter no. PSW-32 / 98-99 (ERO)),

\end{document}